# Far- and near-field photon noise limits to the detectivity of nanometer-thick thermal detectors

**Olivier Merchiers[1], Aapo Varpula[2], Kirsi Tappura[2], Pierre-Olivier Chapuis[1], Mika Prunnila[2]**

[1]INSA de Lyon, CETHIL, UMR5008, 69621 Villeurbanne, France

[2]VTT Technical Research Centre of Finland Ltd, Tietotie 3, FI-02150 Espoo, Finland

==========================================================================

Thermal-radiation detectors such as bolometers —often found as thin, suspended films— are intrinsically limited by their optical absorption properties and by their intrinsic thermal conductive and radiative losses. We analyze the impact of the photon energy exchange between the film and a substrate located close to each other, noticing that the associated near-field exchange has been overlooked and could reduce the detectivity. In addition, we study how the nanometric thickness of the suspended film and its material impact the detectivity of such sensors. It is found that the blackbody-related photon noise limit of the detectivity can be surpassed with thin films. These results emphasize pathways for improvements of thermal-radiation detectors.

==========================================================================

## 1. Introduction

Various applications, from night vision thermal imaging to remote temperature measurements and spectroscopic chemical analysis, use the part of the electromagnetic spectrum where the photon wavelength is longer than visible light, known as the infrared (IR) [1–3]. IR imaging and spectroscopy can be applied, for example, to detect cancerous cells, measure body heat in medicine and sports, monitor processes in industry, and study material changes [4–7]. Thermal-radiation sensors such as bolometers, which transform the radiation-induced temperature rise of an absorber into an electrical signal, provide a cost-effective route for detection in the infrared (IR) and THz spectral ranges [8–11]. A significant advantage is that they can be operated near room temperature contrary to the commonly used quantum detectors that are based on electron-hole pair generation and need cooling to reach their maximal performance, especially, for the small-energy photons of the long-wavelength infrared (LWIR) [9]. Bolometers are typically based on a suspended membrane equipped with an absorber, a construction enabling a strong response to the absorption of photons via heating the membrane. Among recent examples, a nano-thermoelectric bolometer was developed by some of us [12–14] with a specific detectivity that can approach the fundamental limit of the thermal fluctuation noise of thermal detectors [15]. Such a fundamental limit has been defined in the literature based on an opaque bolometer that is coupled with (far-field) blackbody radiation. However, most of the current bolometers involve thin-film structures that are partially transparent and often include back-reflectors [16–18] that are not necessarily located in the thermal far-field. For thin film-detectors, this is specifically the case of the configuration known as the Salisbury screen, where absorption in the thin film is enhanced by placing a reflector at a distance equal to $\lambda/4$, where $\lambda$ is the targeted wavelength, behind the absorber [19-21]. In spite of the sub-wavelength distances, only the far-field effects are traditionally taken into account in the design considerations of such detectors. In addition, it was realized in the last decade that suspended ultrathin metallic films can be excellent absorbers for thermal radiation [22]. For instance, this was studied in [16] for the particular case of TiN films suspended over doped silicon and recently in [23] for oxide materials.

The detectivity of thermal detectors is affected by several noise sources with the main ones being Johnson-Nyquist noise coming from the electrical resistors, the noise from the heat losses suffered by the absorber element and 1/f noise at low frequencies [8,24]. The contributions of the different noise sources depend on the design and the method used for the temperature measurement. As bolometers are usually operated in vacuum, the relevant heat-loss channels are conduction and radiative heat transfer. The thermal conductance associated to heat dissipation by conduction from the bolometer is essentially linked to the actual geometry of the supporting arms, often based on nanofabrication rules, and can be considered as fixed for a selected design. If the area of the detector is increased enough, its contribution becomes smaller than that of the radiative conductance. In this work, we focus on the thermal conductance associated with thermal radiation which includes both the far-field and near-field heat transfer. Especially, we study the impact of the near-field heat transfer on the performance and design rules of a bolometer, i.e., a topic completely overlooked previously despite the typical distances between the absorber and back-reflector essentially involving the thermal near-field range. We focus on the photon noise-based detectivity of thermal-radiation detectors and study systematically the influence of the thickness and position of a thin absorber film above a reflector on the specific detectivity of a bolometer.

The paper is structured as follows. We first introduce the standard formulation of the fundamental detectivity limits. Second, we quickly describe the main equations necessary to describe the near-field radiative heat transfer between a film and the bulk substrate cons. We then detail the main results obtained for a TiN thin film and an aluminum substrate in the case of broadband thermal-radiation detection. It is shown how the heat flux between the substrate and the film impacts the detectivity. We then analyze the same case by changing the material of the absorber. In a second step, we consider spectral and normal illumination, as often the case in spectral detectors. Finally, we analyze the case of a doped silicon back-reflector. The discussion allows suggesting ways for optimizing the structure considering the material properties.

## 2. Detectivity limit associated with radiative heat transfer

Depending on the application, different thermal detector architectures can be considered. When the response time is not a critical parameter, one can choose an optically thick absorber, where the whole incident optical signal is absorbed and no back-reflector is needed. The radiative part of the thermal conductance consists of the radiative heat transfer between the absorber and the surrounding medium, or environment (vacuum). The thermal-radiation losses are then only due to emission into the surroundings.

For applications where the response time is critical such as in thermal imaging, thin absorbers are preferred to reduce the heat capacity. Such a detector architecture is shown in Figure 1. In general, a thin absorber film is not thick enough for absorbing the whole incident optical signal and part of the incoming radiation is lost. In the case of resistive metals, the maximum absorption is found to be 50% when the absorber is impedance matched to a half of the vacuum impedance [25]. Typical absorber materials are metals, semi-metals, semiconductors and dielectrics [26]. To improve the absorption of the signal in an optically-thin absorber a back-reflector is used under the absorber film with a matching cavity in-between (Figure 1). In general, the back-reflector is not a perfect mirror and hence also produces parasitic losses, but they usually stay small. The radiative thermal conductance consists now of two contributions: $G_\text{R}$ for losses towards the vacuum and $G_\text{R,back}$ for the losses towards the back-reflector (see Figure 1). In the case of resistive metals, maximum absorption of 100% can be reached for certain wavelengths when the absorber is impedance-matched with the vacuum impedance. For the back-reflector two types can be distinguished: 1) metallic mirrors, usually made from metals, semi-metals or highly doped semiconductors and 2) distributed Bragg reflectors made from dielectrics or non-absorbing semiconductors [25].

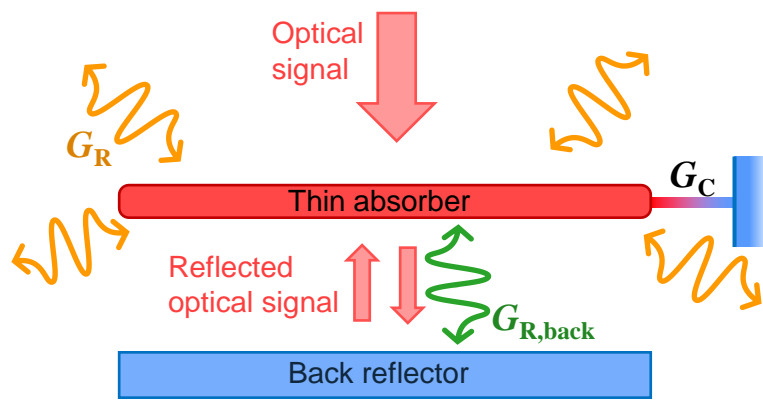

**Figure 1.** Schematic picture of a thermal detector based on optically-thin absorber with a back-reflector. Part of the optical signal is transmitted through the absorber and reflected back to the absorber from the back-reflector. $G_\text{R}$ is the radiative thermal conductance between the absorber and the environment, and $G_\text{R,back} = G_\text{ff} + G_\text{nf}$ is the radiative thermal conductance between the absorber and the back-reflector consisting of the far-field ($G_\text{ff}$) and near-field ($G_\text{nf}$) components. For the sake of completeness, $G_\text{C}$, the conductive thermal conductance associated with the conductive heat transfer along the mechanical supports of the absorber is also shown.

This configuration allows recycling (a part of) the transmitted thermal radiation back towards the absorber. We will see that it also reduces parasitic heat losses towards the environment. The gap between the film and substrate forms an optical cavity and by setting the cavity size to $n\lambda/4$, where $n$ is an odd integer, we optimize the absorption at $\lambda$ [25]. At room temperature, the typical wavelength of thermal radiation is around 10 µm, which sets the cavity size to 2.5 µm for the typical value of $n = 1$. For such small cavity sizes, classical radiative heat transfer does not necessarily apply any longer and one should take the near-field effects into account [22]. In Figure 1, this is included in $G_{R,back} = G_{ff} + G_{nf}$, the radiative thermal conductance between the absorber and the back-reflector consisting of the far-field ($G_{ff}$) and near-field ($G_{nf}$) components. The near-field effects are a result of the tunneling of the evanescent modes located near the substrates and can increase the heat flux between the absorber and the substrate by orders of magnitude compared to the classical value given by the Stefan-Boltzmann law. As a result, $G_{R,back}$ can become large and could in principle deteriorate the detectivity significantly.

Here we go through the standard way (see e.g. [15,24]) to calculate the fundamental detection limit of a thermal detector. Noise-equivalent power (NEP) equals to the input power $P_n$ that produces an output signal with the signal-to-noise ratio (SNR) of unity, $SNR = 1$, at a specified bandwidth $B$, which is often 1 Hz: $\text{NEP} = P_n/\sqrt{B}$. In the case of a thermal detector, the NEP corresponding to the thermal fluctuation noise is given by (cf. [8,10])

$$\text{NEP}_{th} = \frac{\sqrt{4k_B T^2 G}}{\eta}, \qquad (1)$$

where $k_B$ is Boltzmann's constant, $T$ the absolute temperature, $G$ the thermal conductance from the absorber to the environment, and $\eta$ is the absorptance of the film, which indicates the optical efficiency of the absorber. The specific detectivity is defined by

$$D^* = \frac{\sqrt{A}}{\text{NEP}}, \quad (2)$$

where *A* is the absorber area. The specific detectivity $D^*$ corresponding to eq. (1) is given by

$$D^*_{\text{th}} = \frac{\sqrt{A}}{\text{NEP}_{\text{th}}} = \sqrt{\frac{\eta^2 A}{4k_B T^2 G}}, \quad (3)$$

which is valid for all the configurations [15,27].

We now list the typical assumptions made in previous works, which apply to the configuration of Figure 1. If the surroundings are assumed to be at thermal equilibrium (except the radiation to be detected) in the far-field, the radiative part of the thermal conductance is given by the derivative of the Stefan-Boltzmann law with respect to temperature (cf. [27]):

$$G_R = 4A\epsilon\sigma T^3, \quad (4)$$

where $\epsilon$ is the emissivity and $\sigma$ the Stefan-Boltzmann constant. Since emission and absorption coefficients are equal in many cases, Kruse [27] uses $\eta$ in this formula instead of $\epsilon$. Note that such expression assumes implicitly that the absorber is exchanging thermal radiation with a half space only (full space would require the effective area to be 2*A*) and that we deal here implicitly with broadband illumination of the bolometer (see Sec. 5 for the case of spectrally-resolved illumination). Combining eqs. (1-4) with $\epsilon = \eta$ gives

$$\text{NEP}_{\text{th}}(G_R) = \sqrt{\frac{16k_B \sigma A T^5}{\eta}} \quad (5)$$

and

$$D^*_{\text{th}}(G_R) = \sqrt{\frac{\eta}{16k_B \sigma T^5}}. \quad (6)$$

A more rigorous derivation is given in Chapter 2 of Ref. [27], where photon noise is calculated by integrating Planck's spectrum. The result, the background fluctuation noise-limited *D**, is given by

$$D^*_{\text{BF}} = \sqrt{\frac{\eta}{8k_B\sigma(T^5 + T^5_{\text{bg}})}}, \quad (7)$$

where $T_{\text{bg}}$ is the absolute background temperature. In the case where $T = T_{\text{bg}}$, eq. (7) reduces to eq. (6) and $D^*_{\text{BF}} = D^*_{\text{bb}} = 1.81 \cdot 10^{10}$ cm.Hz$^{1/2}$.W$^{-1}$.

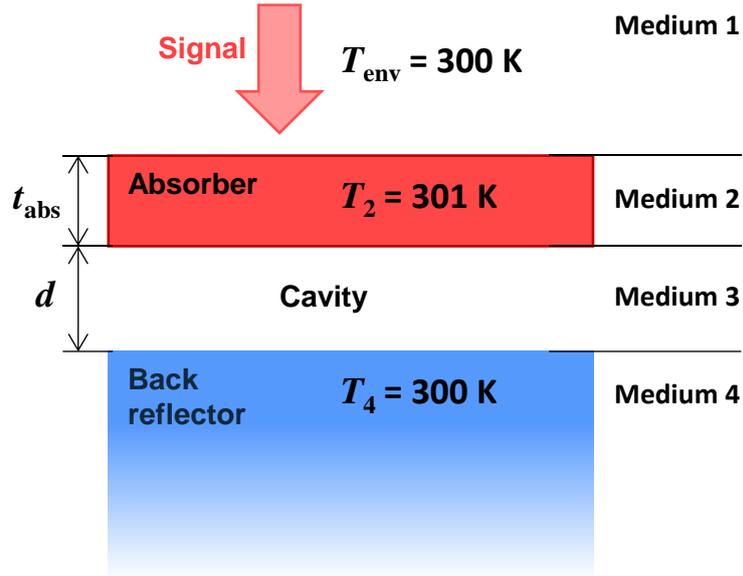

**Figure 2.** Schematic representation of the bolometer model used for the near- and far-field radiative transfer calculations. Unless otherwise stated, all calculations in this work were performed using this structure and temperatures indicated for the different layers and media.

We now consider in more detail a bolometer with an optically-thin absorber as shown Figure 2, assuming that the suspended film is not opaque. The total radiative conductance, which enters in eq. (3), is then:

$$G = G_{\text{tot}} = G_{R,\text{back}} + G_R; \quad G_{R,back} = G_{\text{ff}} + G_{\text{nf}}, \quad (8)$$

where $G_R$ is now a thickness-dependent expression, which cannot be described by the usual surface emissivity, and $G_{R,\text{back}}$ can include both far-field ($G_{\text{ff}}$) and near-field ($G_{\text{nf}}$) contributions due to the exchange between the suspended film and the back-reflector. In the case of a perfect metal $G_{\text{R,back}} = 0$, but even the best noble metals absorb slightly in the far field. In addition, the proximity of the back-reflector to the film can induce a near-field exchange ($G_{\text{nf}}$), which also contributes to a finite value for $G_{\text{R,back}}$.

To account for such near-field interactions, it is needed to resort to Fluctuational Electrodynamics (FED), which is a formulation of Maxwell's equations including a random source term for the current densities [22,28]. This allows reducing the problem of radiative heat transfer to a purely electromagnetic problem. In brief, thermal emission is described in FED as the result of the random thermal motion of the electrons and crystal atoms in the materials. The statistical properties of the random currents are described by the Fluctuation-Dissipation Theorem (FDT) establishing a link between the fluctuations of the current densities and the electromagnetic dissipation expressed through the imaginary part of the permittivity. The bolometer is modeled by a one-dimensional stack of homogeneous layers (Figure 2), which involves the absorber film (medium 2) and the bulk substrate (medium 4) separated by a vacuum gap, or cavity (medium 3). We consider the space above the absorber to be vacuum also (medium 1). Electromagnetic properties of the multilayer stack are completely determined by the layer thicknesses and their complex reflection and transmission coefficients, which depend on their permittivities. The back-reflector is assumed to be aluminum or silicon, modeled as a Drude oscillator:

$$\varepsilon(\omega) = \varepsilon_\infty - \frac{\omega_p^2}{\omega(\omega + i\gamma)}, \qquad (9)$$

where $\omega_p$ is the plasma frequency and $\gamma$ the damping rate [29]. For the absorber material, we will consider different metals such as TiW, TiN, Ti, W, and Au. All of these are modeled using a Drude dispersion model with parameters obtained from literature [30] or from inhouse measurements. In Table 1 we provide the values of the parameters for all the materials considered in this work.

**Table 1**. Values of the parameters of the Drude model for the materials considered in this study. $m_0$ is the mass of the electron, $e$ the elementary charge, $n$ the charge carrier density, $\rho$ the electrical resisitivity and $\varepsilon_0$ the vacuum permittivity.

| Material | $\varepsilon_\infty$ | $\omega_p$ (rad.s$^{-1}$) | $\gamma$ (s$^{-1}$) |
|---|---|---|---|
| Al[a] | 1.0 | $2.24 \times 10^{16}$ | $1.22 \times 10^{14}$ |
| Au[a] | 1.0 | $1.37 \times 10^{16}$ | $5.32 \times 10^{13}$ |
| Ti[a] | 1.0 | $3.82 \times 10^{15}$ | $719.6 \times 10^{11}$ |
| W[b] | 6.0 | $6.2 \times 10^{15}$ | $2.6 \times 10^{14}$ |
| TiN[c] | 1.0 | $7.55 \times 10^{15}$ | $6.67 \times 10^{14}$ |
| TiW[c] | 1.0 | $7.52 \times 10^{15}$ | $1.0 \times 10^{15}$ |
| Si (n doped)[d] | 11.7 | $\sqrt{\dfrac{ne^2}{0.27 m_0 \varepsilon_0}}$ | $\dfrac{ne^2 \rho}{0.27 m_0}$ |
| Si (p doped)[d] | 11.7 | $\sqrt{\dfrac{ne^2}{0.34 m_0 \varepsilon_0}}$ | $\dfrac{ne^2 \rho}{0.34 m_0}$ |

[a]Ref. [28], [b]From inhouse measurements, [c]Ref. [12], [d]Ref. [31]

The net heat flux between the absorber film and the bulk substrate (back-reflector) can be decomposed in a propagative (far-field) and an evanescent (near-field) contributions [22]:

$$q_{\text{prop}} = \int_0^\infty \frac{d\omega}{2\pi} [\Theta(\omega, T_2) - \Theta(\omega, T_4)] \sum_{\text{TE,TM}} \int_0^{\frac{\omega}{c}} \frac{KdK}{2\pi} \frac{\left(1 - \left|R_{32}^j\right|^2 - \left|t_{21}^j\right|^2\right)\left(1 - \left|R_{34}^j\right|^2\right)}{\left|1 - R_{32}^j R_{34}^j e^{2ik_{z3}d}\right|^2},$$

$$q_{\text{evan}} = \int_0^\infty \frac{d\omega}{2\pi} [\Theta(\omega, T_2) - \Theta(\omega, T_4)] \sum_{\text{TE,TM}} \int_{\frac{\omega}{c}}^\infty \frac{KdK}{2\pi} \frac{4\text{Im}\left(R_{32}^j\right)\text{Im}\left(R_{34}^j\right)e^{-2\text{Im}(k_{z3})d}}{\left|1 - R_{32}^j R_{34}^j e^{2ik_{z3}d}\right|^2},$$

(10)

and the radiative conductance can be obtained from

$$G_{\text{R,back}} = \lim_{\Delta T \to 0} \frac{S \cdot (q_{\text{evan}} + q_{\text{prop}})}{\Delta T}, \tag{11}$$

where $S$ is the area of the detector. In all our calculations the absorber layer temperature is fixed at 301 K, and the vacuum environment and back-reflector at 300 K (see Fig. 2). In expressions (10), $\Theta(\omega, T_i)$ is the mean energy of a Planck oscillator in thermal equilibrium at angular frequency $\omega$ and temperature $T_i$, with $T_2$ and $T_4$ the temperatures of the film and the substrate. $c$ is the speed of light in vacuum. $R_{kl}^j$ is the reflection coefficient for a plane wave incident on medium $l$ incoming from medium $k$ with polarization $j$, whereas $t_{21}^j$ is the total transmission coefficient of medium 2 towards medium 1 (including multireflections). $d$ is the size of the vacuum gap separating both bodies and finally $k_{zi}$ is the $z$-component of the wavevector in medium $i$ and $K$ is the component of the wavevector parallel to the interface. For the propagative component of the heat flux, we account for the semi-transparency of medium 2 since it is optically thin. The terms $1 - |R_{32}^j|^2 - |t_{21}^j|^2$ and $1 - |R_{34}^j|^2$ are the (size and polarization-dependent) monochromatic directional emissivities of medium 2 and 4, respectively, and their presence in the heat flux equation are a signature of Kirchhoff's reciprocity law. In the first component integration over the parallel wavevector $K = \frac{\omega}{c}\sin\theta$ is performed from 0 to $\omega/c$ which corresponds to propagating modes with an angle of incidence on the interfaces $\theta$ between 0 and $\pi/2$. The evanescent contribution is obtained by integrating from $\omega/c$ to $+\infty$. For such values of $K$, $k_{z3}$ is purely imaginary describing evanescent modes originating from total internal reflection or surface modes including surface phonon-polaritons [22]. For gap sizes larger than Wien's wavelength, the evanescent modes are negligible, and the heat flux is purely the result of propagative modes and does not depend on the gap size thereby recovering Stefan-Boltzmann's law. In the near-field region, however, it depends strongly on the gap size.

The contribution $G_R$ giving the heat losses towards the upper environment at 300 K is also computed rigorously in the frame of FED. In this case, however, no near-field effects occur and only propagating contributions are accounted for. We place a blackbody (assimilated numerically to a permittivity $\varepsilon = 1 + 10^{-5}i$ ) at temperature of 300 K at a distance of 1 cm from the absorber film (assimilated numerically to an arbitrarily-large distance). The heat flux between the blackbody and absorber film is given by

$$q_{\text{bb-film}} = \int_0^\infty \frac{d\omega}{2\pi} [\Theta(\omega, T_2) - \Theta(\omega, T_{\text{bb}})] \sum_{\text{TE,TM}} \tau_{\text{bb-2}}(\omega), \qquad (12)$$

where $T_{\text{bb}}$ is the temperature of the blackbody and $\tau_{\text{bb-2}}$ the transmission function taking into account the interaction between the absorber film and the back-reflector rigorously using the S-matrix formalism described in [28]. By applying the definition given by Eq. (11), we can similarly obtain the conductance $G_{\text{R}}$. The same calculation can be used to obtain numerically the total hemispherical absorptivity $\eta$ of the thin film given by

$$\eta = \frac{1}{\sigma T_{\text{bb}}^4} \int_0^{+\infty} d\omega\, \Theta(\omega, T_{\text{bb}}) \sum_{\text{TE,TM}} \tau_{\text{bb-2}}(\omega). \qquad (13)$$

This expression is size dependent and involves contributions by the back-reflector.

## 3. Broadband detectivity for the TiN absorber/Al back-reflector case

In a realistic device, the detectivity value is the result of different noise and loss mechanisms. Here we exclusively focus on the contribution of the heat losses of the absorber towards the environment and substrate. This depends on the geometry of the bolometer and its constituent materials. We consider an absorber made of TiN, often used in bolometer design due to its high absorptivity [16], and analyse the effect of the absorber thickness and cavity size. For a highly-reflective substrate, we assume aluminium, which is expected to achieve the best performances. In this section, broadband absorption is considered.

### 3.1 Impact of cavity size and film thickness

Fig. 3 shows the results of radiative heat losses and detectivity for a TiN absorber and an aluminum back-reflector. The thickness of the absorber is varied from 4 to 50 nm and the gap size from 10 nm up to 100 µm. For the detectivity calculations we set the absorptivity of the absorber to $\eta = 1$ (see Eq. (3)) to focus first on the radiative losses. The optical properties of TiN and Al are described using the Drude model (Eq. (9)) using the values of bulk media, so a potential thickness-dependent correction of the Drude losses is not included.

Fig. 3(a) shows the different components of the heat flux (towards environment, substrate and the sum of both) for an absorber thickness of 12 nm. For gap sizes larger than 5 micrometers, the heat flux towards the environment remains constant and leads the losses, as initially expected from a back-reflector. However, since the flux between the absorber layer and substrate accounts for both the near- and far-field contributions, we find a characteristic three-orders-of-magnitude increase compared to the far-field value given by Stefan-Boltzmann's law for small gap sizes (see blue curve in Fig. 3(a)). Interestingly, the flux towards the environment also changes for the gap sizes below a few micrometers. A small local maximum of the total flux lost is observed, which is a signature of interference occurring in the cavity, as expected from the Salisbury configuration but which seems to have a negative effect here.

In Fig. 3(b) we represent the total radiative heat losses towards the environment and the substrate as a function of the size of the cavity for different film thicknesses. For large thicknesses, the film is almost opaque and the behavior follows that of a semi-infinite bulk material of TiN at large gap distances. It is worth noting that the thinnest films seem to be the strongest emitters, in contrast to common sense that a larger volume allows to emit better (think of Kirchhoff's law that states that a good emitter is a good absorber). It has been shown that single metallic layers show strong absorption compared to bulk metals [22], which manifests here. At first sight, one would therefore discard such thin absorbers. However, for the small thicknesses (semitransparent films), we observe interference effects in the 100 nm to a few micrometers range, with a minimum flux for gap sizes of a few hundreds of nanometers. This minimum is of interest.

We turn now to Figs. 3(c-d) in which we represent the specific detectivity, in the units of black body detectivity, as a function of the gap size. According to the definition of detectivity the behaviors are the reciprocals of those of the heat losses. We point out that the total detectivity is given by the lower limit (red curve of Fig. 3(c)) of both near- and far-field detectivities. This inverse relation between heat flux and detectivity leads to a detectivity strongly dependent on the gap size in the near-field, and almost constant for the far-field losses. In particular, a detectivity peak is observed in the transition zone between the near-field and far-field regions, which is easy to understand in the light of the radiative flux evolution presented in Fig. 3(a). The position and amplitude of this maximum is sensitive to the shapes of the detectivity curves in that region and will change with layer thickness and absorber material. We note that the position of the maximum occurs for a cavity size which is smaller than the commonly used 2.5 µm (or $\lambda/4$ for $\lambda$ = 10 µm) at room temperature [31].

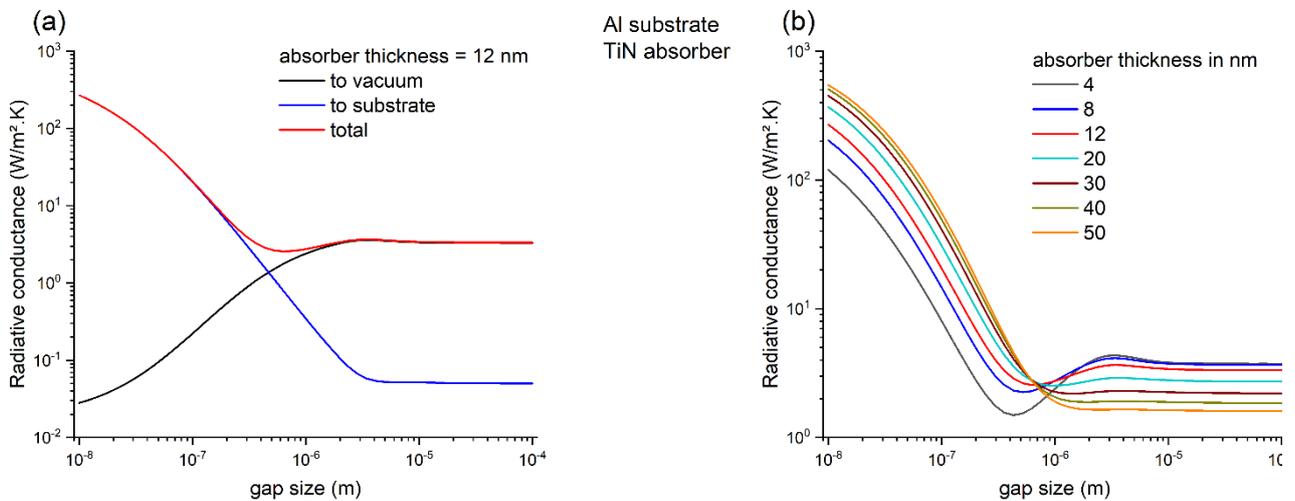

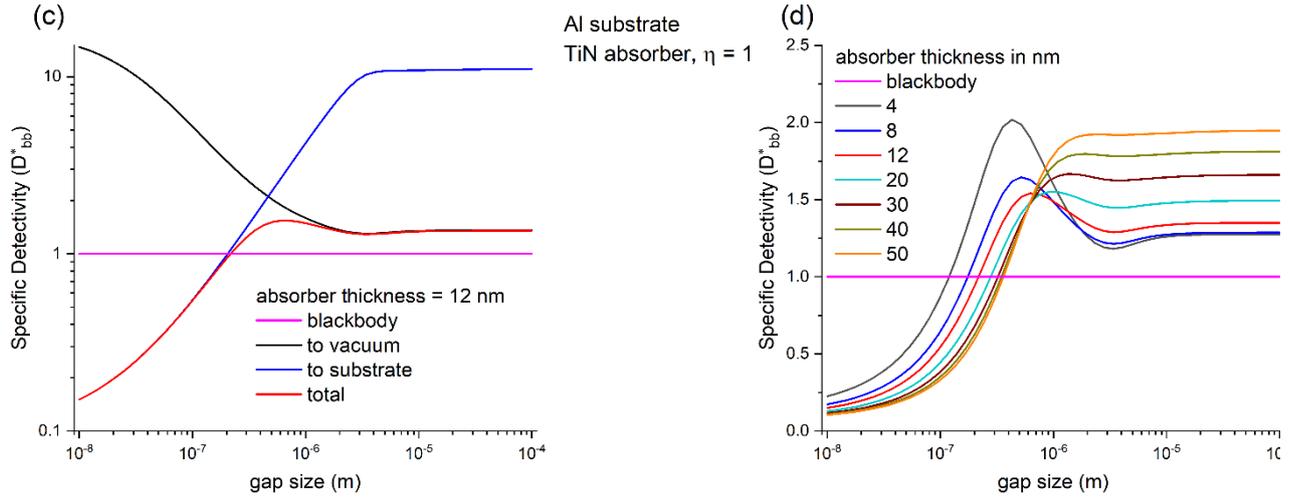

**Figure 3.** Results for homogeneous TiN films with finite thicknesses over an aluminum substrate as a function of the size of the vacuum gap. (a) Radiative conductance for a 12-nm thick absorber, between the absorber and environment (black curve), absorber and substrate (blue curve) and the total heat flux (red curve). (b) Radiative conductance towards the environment and the back-reflector for different absorber thicknesses. (c-d) Specific detectivity in units of $D_{bb}^\star = 1.81 \cdot 10^{10}$ cm Hz$^{1/2}$ W$^{-1}$ as a function of the gap size associated with (a-b). The purple curve gives the blackbody value. Note that the absorptivity of the absorber has been artificially set to unity ($\eta = 1$) in all the cases.

Note that the calculations do not include any correction for the boundary scattering by the electrons due to the small layer thickness. Including this effect would increase the imaginary part of the TiN permittivity and, as a consequence, increase the thermal emission by the absorber layer leading to a change in the amplitude and shape of the detectivity peak. For all thicknesses, the specific detectivity computed here can be above the blackbody one. However, the real absorptivity was not included in this section.

### 3.2 Including broadband absorptivity

All the results presented above were obtained for the total hemispherical absorptivity equal to 1 in order to focus on the loss term. However, this typically leads to an overestimation of the detectivity (see Eq. (3)) and neglects the gap size dependence. We now compute the correct TiN total hemispherical absorptivity from the S-matrix formalism. As shown in Fig. 4(a), the total absorptivity (eq. (13)) is always below 1. The highest absorptivity (≈ 0.7) is obtained for a 4 nm thick TiN absorber and occurs for a gap size of approximately 4 µm. We find again that thin metallic films absorb very well. The impact of including a real material as the back-reflector and including a finite thickness for the absorber metal seems however to shift the typical value of the Salisbury screen (∼2.5 µm) to a larger value. We show in Fig. 4(b) the specific detectivity as a function of gap size for different absorber thicknesses. Comparing the black curve (4 nm thickness) in Fig. 3(d) with the black curve of Fig. 4(b), we see that in the first figure the detectivity reaches two times the blackbody value for a gap size of 300 nm while in the second (the accurate one) the maximum detectivity value drops approximately to 0.8 and moves to the value of 4 µm on the gap size axis. Indeed, the maximum detectivity is now dictated by the maximum absorptivity. The large peak visible in Fig. 3(d) has completely disappeared in Fig. 4(d) because at 400 nm the absorptivity barely reaches a value of 0.1.

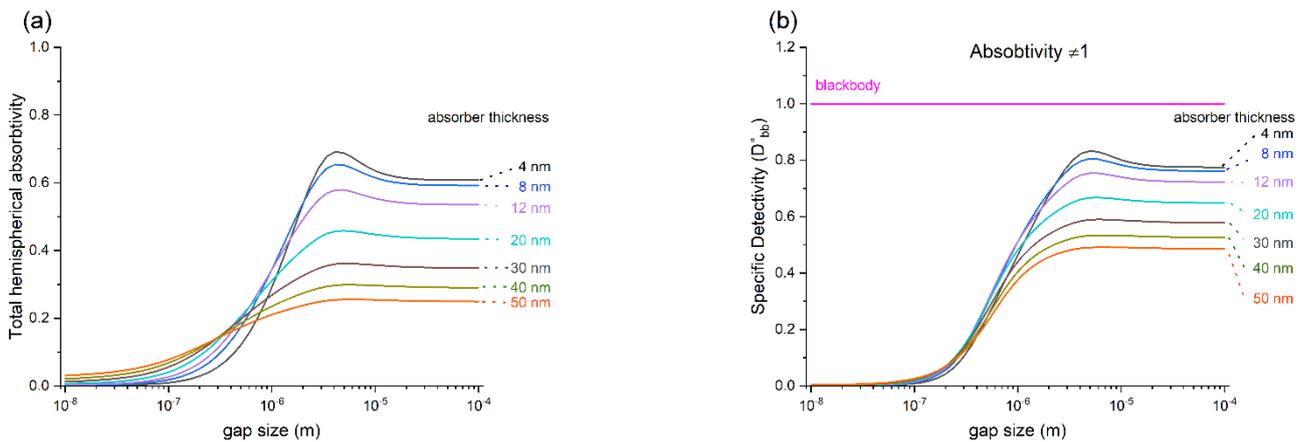

**Figure 4.** (a) Total hemispherical absorptivity as a function of gap size for different TiN thicknesses. (b) Specific detectivity as a function of gap size for different TiN absorber thicknesses. The purple line shows the blackbody specific detectivity.

## 4. Absorber material effects on broadband detectivity

We have shown in the previous sections that the detectivity peak is sensitive to the absorber layer thickness but also to the detailed behavior of the near- and far-field heat fluxes in the vicinity of the maximum. This behavior is known to be strongly material dependent and hence we expect the maximum to change for different materials. In this novel step, we extend the analysis to different materials to find the best candidates for the absorber.

The different metals studied for the absorber, besides TiN, are TiW, Ti, W and Au. Their optical properties are modeled with parameters listed in Table 1. The results of the detectivity as a function of the cavity size for those materials are presented in Fig. 5(a) for an absorptivity $\eta = 1$ similarly as in Sec. 3.1. We see the highest detectivity (in units of $D_{bb}^{\star}$) for gold reaching values close to 3.5. The lowest detectivity is obtained for TiN. This is easily explained by the fact that the highest detectivities are obtained for the metals with highest reflectivities such as gold, titanium and tungsten. High reflectivity, however, implies small absorption by the absorber and, hence, small temperature increases and poor detector output signals. It is therefore essential to add the absorptivity of the material to obtain a realistic estimate of the detectivity. The total (broadband) absorptivity was included in Fig. 5(b). We immediately see that the trend compared to Fig. 5(a) is completely reverted, and TiN together with TiW become the best candidates. As for TiN, their detectivities have been divided by 2 compared to the values in Fig. 5(a). For gold, Ti and W the reduction factors range between 4 (Ti, W) and 50 (Au). Again, while all the far-field detectivities are larger than the black body limit in Fig. 5(a), including the absorptivity brings all of them below the blackbody limit.

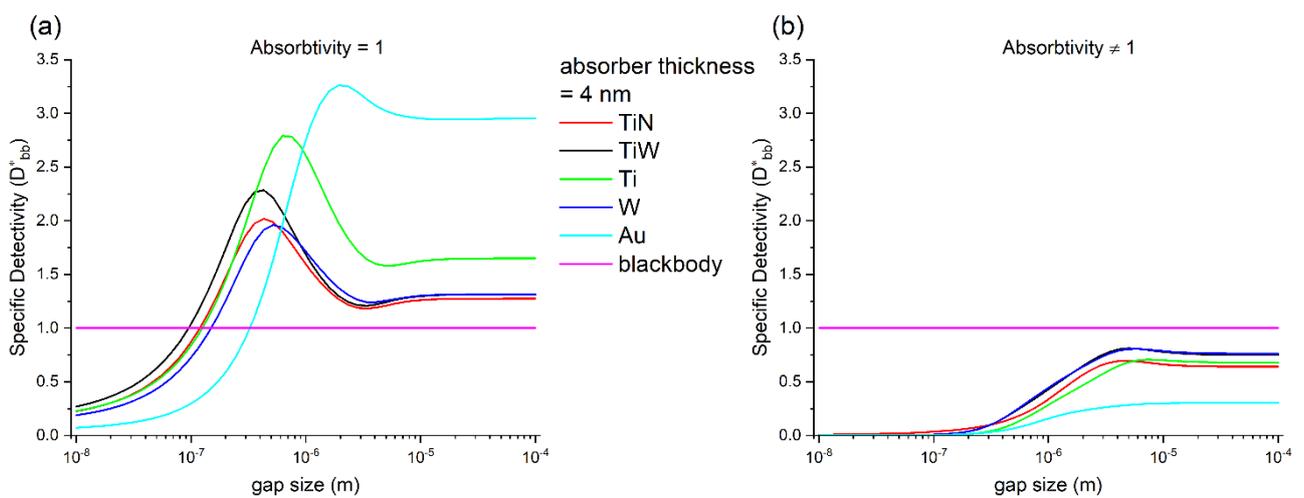

**Figure 5**. Relative specific detectivities as a function of cavity size for different metals: TiN, TiW, Ti, W and Au. All the absorbers have a thickness of 4 nm. (a) The absorptivity is assumed equal to 1 ($\eta = 1$). (b) Actual absorptivity of the absorber layer is included.

We note that the maxima of broadband detectivity are reached for cavity sizes around 3 to 4 µm, i.e. near-field effects do not seem to impact strongly the detectivity at first sight. We analyze their positions in more detail in Fig. 6 as they are expected to depend on the gap size, absorber thickness and its material properties. In Fig. 6(a) we see how including the total (broadband) absorptivity changes the behavior of the peak position radically. For $\eta = 1$, all detectivities are larger than the blackbody one, showing a local minimum around 10 nm thickness except for gold for which the behavior is purely monotonic. This local minimum corresponds with the transition observed in Fig. 3(d) between the large detectivity maximum for small absorber thicknesses (4 to 12 nm) and the flat detectivity for large thicknesses (20 to 50 nm). Including the realistic absorptivity leads to a slow decrease of the maximum value, matching with the behavior of Fig. 4(a) since absorptivity reduces with increasing thickness. The behavior of gold now matches the trends of other metals, even though the value of the maximum is almost one order of magnitude smaller.

When looking at the optimum gap size as a function of thickness (Fig. 6(b)), both sets of curves for $\eta = 1$ and $\eta \neq 1$ show a monotonic growth albeit starting from different gap sizes. TiN and TiW for $\eta = 1$ show a jump for a thickness of 40 nm. For those thicknesses the maximum is actually a plateau and the maximum position is hard to pinpoint. This jump does not occur when $\eta \neq 1$, except for gold at 50 nm thickness, since the maximum is mostly determined by the absorptivity maximum.

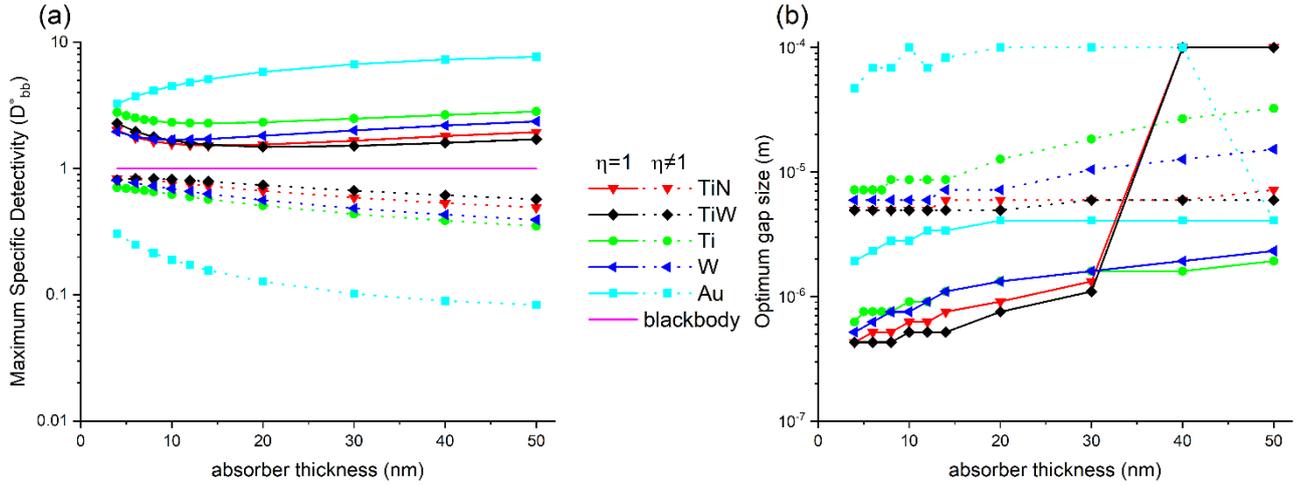

**Figure 6.** Evolution of the maximum detectivity value and position as a function of gap size, absorber thickness and material properties. Solid lines give the results for absorptivity set to $\eta = 1$, dotted lines include the computed value of the absorptivity. (a) Maximum value of the specific detectivity as a function of absorber thickness for different materials. (b) Optimum gap size corresponding with the maximum detectivity value as a function of absorber thickness for the considered materials.

## 5. Spectral absorptivity and detectivity

While in Sections 3-4 we have computed the total hemispherical absorptivity (integrated over all incident directions and wavelengths), we now investigate the spectral absorptivity at normal incidence $\eta(\lambda)$, which provides data for a different type of bolometers, which is spectrally resolved. This could be set in practice by adding a filter before the detector. The detectivity is then given by (cf. eq.(3))

$$D_{\text{th}}^*(\lambda) = \sqrt{\frac{\eta(\lambda)^2 A}{4k_{\text{B}}T^2 G}} \;, \qquad (14)$$

where $G$ is the total radiative conductance integrated over all frequencies and is exactly the same quantity we used in all the previous calculations. Fig. 7(a) provides the relative specific detectivity as a function of the incident wavelength and gap size. The substrate is still aluminum and the absorber TiN with 4 nm thickness. The map shows the expected interference maxima and minima typical of a Fabry-Pérot cavity. As expected, maxima and minima positions move to larger wavelengths for larger gap sizes. Below 100 nm gap sizes, the detectivity is nearly zero and shows up as a black band which extends towards larger gap sizes when the wavelength increases following the behavior of Farby-Pérot modes. We remind that the reason for such low detectivity is because, in this region, radiative heat losses are dominated by near-field effects which can be up to three orders of magnitude larger than the losses towards the upper medium. To gain a better understanding we take a section of this map at a wavelength of 10 µm (Wien's wavelength at 300 K). This is represented by the red curve in Fig. 7(b). We also added the detectivity with $\eta = 1$ (black curve) and the detectivity including the total absorbtivity $\eta \neq 1$. As expected, the black curve acts as the envelope of the red curve since the latter one is just obtained by multiplying the detectivity with $\eta = 1$ by the spectral absorptivity at normal incidence. We see that the detectivity including the total absorptivity $\eta \neq 1$ (blue curve) barely reaches a value of 0.7, while the red curve goes up to 1.2. This difference is simply due to the fact the former is integrated over all wavelengths and directions, leading to lower values. These results show that for monochromatic normally incident radiation, the specific detectivity can be larger than for broadband blackbody radiation. Interestingly, the fact that the spectral data (red curve) do not reach the region where the black curve is at its maximum can lead to assume that there is still a phase space for improving the specific detectivity: it would be needed to find some configuration where spectral absorptivity is higher for smaller gap sizes.

(a)            (b)

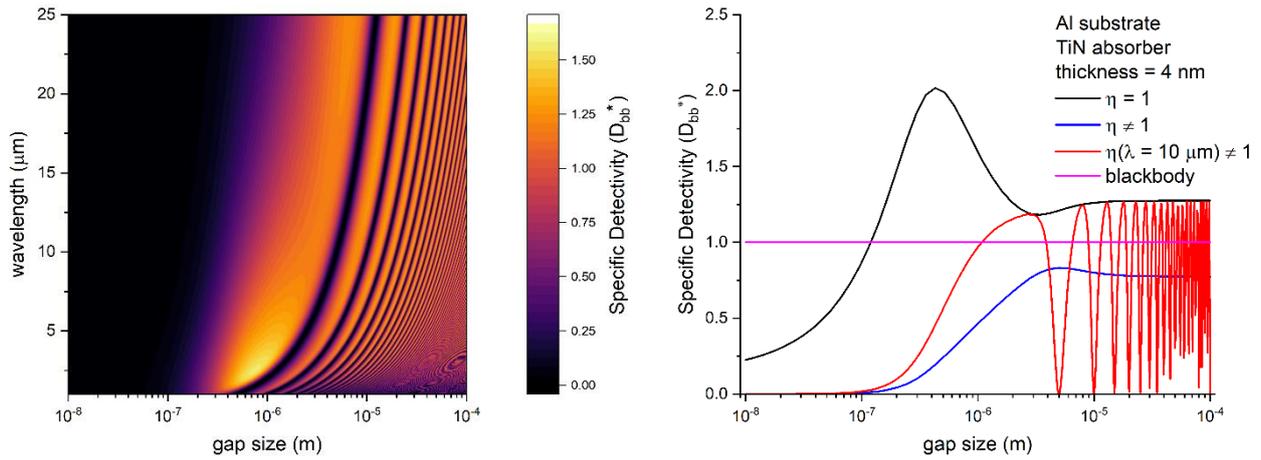

**Figure 7.** Relative specific detectivity for a TiN absorber of 4 nm in thickness above an aluminium substrate (a) including spectral absorptivity at normal incidence. The color scale is the spectral detectivity normalized by the blackbody value $D^*_{bb}$ = 1.81·10$^{10}$ cmHz$^{1/2}$/W. (b) Relative detectivity as a function of gap size with the spectral absorptivity integrated over all incident directions and wavelengths (blue curve) and for wavelength = 10 μm with the corresponding spectral absorptivity at the normal incidence (red curve). The black curve is the total relative detectivity with absorptivity equal to 1.

## 6. Impact of back-reflector material

The aim of a Salisbury screen is to suppress reflection towards the outer medium and hence improve absorptivity and detectivity inside the thin film. To have a perfect $\lambda/4$ cavity, however, a perfect reflector is needed to produce the highest quality factor. In all previous calculations, we have used highly-reflective aluminum substrate for the back-reflector. When including near-field effects, a good back-reflector should have the added benefit of reducing near-field parasitic losses towards the same reflector. Since near-field is intrinsically connected to losses (see the imaginary parts in eq. (10)), it would be needed to minimize the losses of the material of the back-reflector. In practice all real metals are not perfect reflectors and have significant losses. Other types of metals could therefore be considered. Another option is to consider highlydoped silicon substrates, which have a metallic character and are typically chosen for fabrication reasons. While being heavily doped, such substrates unfortunately do not reach as high reflectivity as aluminum. However, due to their widespread use we analyze their impact on detectivity. Here we revert to the case of broadband illumination of the detector.

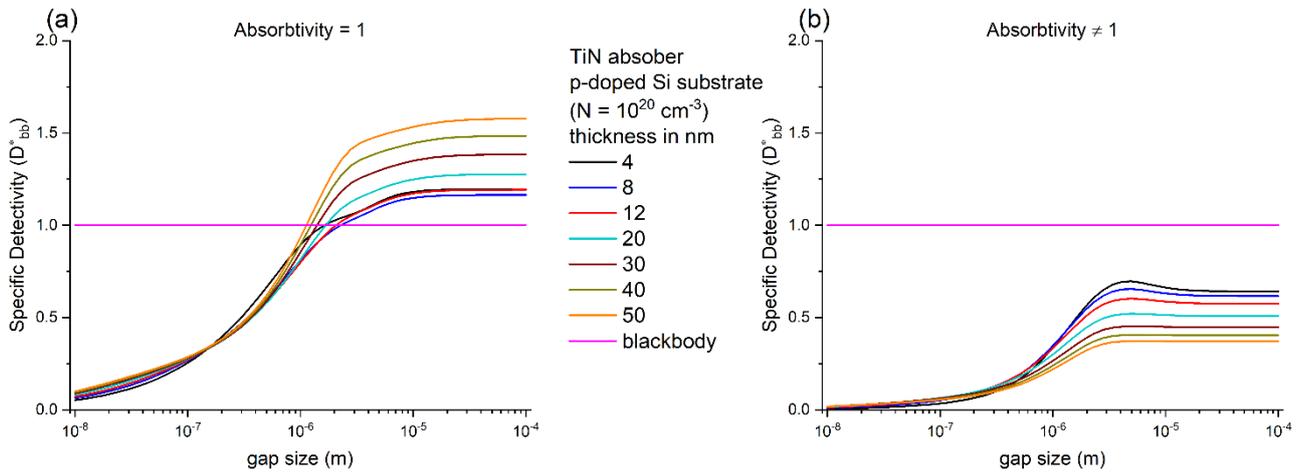

**Figure 8.** Specific detectivity as a function of gap size for different thicknesses of a TiN absorber for the case of broadband illumination. The back-reflector is made from a highly p-doped silicon with the doping level of $10^{20}$ cm$^{-3}$. The results are expressed in units of blackbody specific detectivity ($D^*_{bb}$). (a) Results with absorptivity set to 1 and (b) account for the total absorptivity integrated over wavelengths and directions.

Fig. 8(a) shows the detectivity where we have replaced the aluminum back-reflector with a doped silicon one. Compared with Fig. 3(d), we notice immediately the absence of a strong peak around 400 nm gap size. The detectivity for a silicon back-reflector only reaches a value 1.5 and the slope is smaller than for the aluminum case. This result of a larger near-field flux between absorber and silicon back-reflector is mainly due to the larger losses of doped silicon. For gap sizes larger than 1 µm, the detectivity increases with absorber thickness as expected, however the largest values remain below the ones obtained for the aluminum back-reflector.

When we include the total absorptivity (Fig. 8(b)), the same trend as in Fig. 4(b) is observed, i.e., the thicknesses with the largest detectivity with $\eta = 1$, have now the lowest one and the blackbody limit can no longer be overcome. The reduction for a 4 nm absorber is of the order of 13 % and for the 50 nm thick absorber it is close to 25 %. This observed reduction when including $\eta \neq 1$ is much smaller than that for the $\eta = 1$ case where in the worst case the reduction goes up to 40 % for the 4 nm thick absorber. This indicates that the absorptivity, even though it depends on the material of the back-reflector, is the dominating factor estimating the performances of the device for broadband illumination of the bolometer.

## 7. Conclusions

We have analysed the relative specific detectivity for a simplified model of a bolometer consisting of an absorber element which is a suspended flat homogeneous film with a finite thickness and a substrate acting as the back-reflector. For the absorber, we have considered TiN, TiW, Ti, W and Au and for the back-reflector we used both aluminium and doped silicon. To estimate the detectivity we have computed the parasitic heat losses towards the environment but also towards the back-reflector. We have confirmed numerically that ultrathin films seem to be attractive for these applications. As typical bolometer cavities use a $\lambda/4$ architecture to minimize reflection towards the environment, the distance separating the absorber from the reflector is around 2.5 µm at room temperature and for such distances, near-field thermal transfer can become a significant loss mechanism in addition to macroscopic thermal radiation predicted by Stefan-Boltzmann's law. We have shown that the near-field flux reduces strongly the detectivity for gap sizes smaller than 1 µm for all considered metals, but that above a 1 µm gap separation the heat flux towards the environment dominates and a plateau is reached for the detectivity. Analysing the losses (setting the absorptivity to unity), the best reflectors give the highest detectivities as expected. However, by including the absorptivity in the case of broadband illumination, we have shown that of all the absorber materials considered TiN and TiW are indeed the best candidates. By computing the detectivity using the monochromatic absorptivity at normal incidence (case of a spectrally-selective bolometer), we have found that the blackbody detectivity can be overcome for certain gap separations. Finally, we have looked at the effect of the back-reflector material. By replacing aluminum by highly doped silicon, we showed a decrease of the detectivity between 13% and 25% percent depending on the absorber thickness.

All these results indicate that the blackbody detectivity limit is not the ultimate limit and that it could be broken, especially in the case of a spectrally-selective bolometer. In particular, the detectivity could be enhanced if high absorptivities were achieved for cavity sizes associated with a dip in radiative losses, which include far and near field contributions, at a few hundreds of nanometers. Since the structures considered here have been relatively simple and more complex nanophotonic ones could be considered, this goal seems reachable.

# Acknowledgements

This work has been financially supported by European Union Future and Emerging Technologies (FET) Open under Horizon 2020 program (Grant Agreement No. 766853, project EFINED), by Business Finland co-innovation projects RaPtor and HigPIg (Nos. 6030/31/2018 and 4380/31/2023), and the Academy of Finland (Grant No. 342586).

# References

[1]  G. Gerlach and H. Budzier, *Thermal Infrared Sensors: Theory, Optimisation and Practice* (Wiley, 2011).

[2]  D. Popa and F. Udrea, Towards Integrated Mid-Infrared Gas Sensors, Sensors **19**, 9 (2019).

[3]  F. Baldini, A. D'Amico, C. D. Natale, P. Siciliano, R. Seeber, L. D. Stefano, R. Bizzarri, and B. Ando, editors, *Sensors: Proceedings of the First National Conference on Sensors, Rome 15-17 February, 2012* (Springer-Verlag, New York, 2014).

[4] J. I. P. Quesada, editor, *Application of Infrared Thermography in Sports Science* (Springer International Publishing, 2017).

[5] G. J. Tattersall, Infrared thermography: A non-invasive window into thermal physiology, Comp. Biochem. Physiol., Part A Mol. Integr. Physiol. **202**, 78 (2016).

[6] D. Landgrebe, C. Haake, T. Höpfner, S. Beutel, B. Hitzmann, T. Scheper, M. Rhiel, and K. F. Reardon, On-line infrared spectroscopy for bioprocess monitoring, Appl. Microbiol. Biotechnol. **88**, 11 (2010).

[7] M.-W. Chen, S. You, K. S. Suslick, and D. D. Dlott, Hot spots in energetic materials generated by infrared and ultrasound, detected by thermal imaging microscopy, Rev Sci Instrum **85**, 023705 (2014).

[8] P. L. Richards, Bolometers for infrared and millimeter waves, Journal of Applied Physics **76**, 1 (1994).

[9] A. Rogalski, *Infrared Detectors* (CRC Press, 2000).

[10] U. Dillner, E. Kessler, and H.-G. Meyer, Figures of merit of thermoelectric and bolometric thermal radiation sensors, J. Sens. Sens. Syst. **2**, 85 (2013).

[11] C. Zhang, E. K. Yalavarthi, M. Giroux, W. Cui, M. Stephan, A. Maleki, A. Weck, J.-M. Ménard, and R. St-Gelais, *High Detectivity Terahertz Radiation Sensing Using Frequency-Noise-Optimized Nanomechanical Resonators*, arXiv:2401.16503.

[12] A. Varpula, K. Tappura, J. Tiira, K. Grigoras, O.-P. Kilpi, K. Sovanto, J. Ahopelto, and M. Prunnila, Nano-thermoelectric infrared bolometers, APL Photonics **6**, 036111 (2021).

[13] A. Varpula, A. Murros, K. Sovanto, A. Rantala, D. Gomes-Martins, K. Tappura, J. Tiira, and M. Prunnila, *Uncooled Nano-Thermoelectric Bolometers for Infrared Imaging and Sensing*, in *Optical Components and Materials XX*, edited by M. J. Digonnet and S. Jiang (SPIE, San Francisco, United States, 2023), p. 33.

[14] A. Murros, K. Sovanto, J. Tiira, K. Tappura, M. Prunnila, and A. Varpula, Infrared bolometers based on 40-nm-Thick Nano-Thermoelectric silicon membranes, Infrared Physics & Technology **145**, 105720 (2025).

[15] P. W. Kruse, A comparison of the limits to the performance of thermal and photon detector imaging arrays, Infrared Physics & Technology **36**, 869 (1995).

[16] T. S. Luk, G. Xu, W. Ross, J. N. Nogan, E. A. Scott, S. Ivanov, O. Niculescu, O. Mitrofanov, and C. T. Harris, Maximal absorption in ultrathin TiN films for microbolometer applications, Applied Physics Letters (2023).

[17] Y. Ra'di, C. R. Simovski, and S. A. Tretyakov, Thin Perfect Absorbers for Electromagnetic Waves: Theory, Design, and Realizations, Phys. Rev. Appl. **3**, 037001 (2015).

[18] C. Chen, C. Li, S. Min, Q. Guo, Z. Xia, D. Liu, Z. Ma, and F. Xia, Ultrafast Silicon Nanomembrane Microbolometer for Long-Wavelength Infrared Light Detection, Nano Lett. **21**, 8385 (2021).

[19] B. Chambers, Optimum design of a Salisbury screen radar absorber, Electronics Letters **30**, 1353 (1994).

[20] Zhongxiang Shen and Hang Wang, *On the Optimum Design of a Thin Absorbing Screen*, in *2007 IEEE Antennas and Propagation Society International Symposium* (IEEE, Honolulu, HI, 2007), pp. 6039–6042.

[21] W. W. Salisbury, *Absorbent Body for Electromagnetic Waves*, U. S. Patent 2,599,944 (10 June 1952).

[22] M. Francoeur, M. P. Mengüç, and R. Vaillon, Spectral tuning of near-field radiative heat flux between two thin silicon carbide films, J. Phys. D: Appl. Phys. **43**, 075501 (2010).

[23] X. Zhang et al., Atomic lift-off of epitaxial membranes for cooling-free infrared detection, Nature **641**, 98 (2025).

[24] P. W. Kruse and D. D. Skatrud, *Uncooled Infrared Imaging Arrays and Systems*, Vol. 47 (Academic Press, New York, U.S.A., 1997).

[25] J. J. Talghader, A. S. Gawarikar, and R. P. Shea, Spectral selectivity in infrared thermal detection, Light Sci Appl **1**, e24 (2012).

[26] R. Ambrosio, M. Moreno, J. Mireles Jr., A. Torres, A. Kosarev, and A. Heredia, An overview of uncooled infrared sensors technology based on amorphous silicon and silicon germanium alloys, Physica Status Solidi c **7**, 1180 (2010).

[27] P. W. Kruse, *Uncooled Thermal Imaging Arrays, Systems and Applications* (SPIE Press, Bellingham, Wash, 2001).

[28] S. Edalatpour and M. Francoeur, Size effect on the emissivity of thin films, Journal of Quantitative Spectroscopy and Radiative Transfer **118**, 75 (2013).

[29] John David Jackson, *Classical Electrodynamics* (1998).

[30] M. A. Ordal, R. J. Bell, R. W. Alexander, L. L. Long, and M. R. Querry, Optical properties of fourteen metals in the infrared and far infrared: Al, Co, Cu, Au, Fe, Pb, Mo, Ni, Pd, Pt, Ag, Ti, V, and W., Appl. Opt., AO **24**, 4493 (1985).

[31] Han Ye and Wenquan Che, *Analysis and Design of a Salisbury Screen Absorber with High Impedance Ground Plane*, in *2013 Cross Strait Quad-Regional Radio Science and Wireless Technology Conference* (IEEE, Chengdu, China, 2013), pp. 325–328.